\documentclass[a4paper,11pt]{article}
\usepackage{pos}

\title{Emerging Topics for Hard Probes}

\author*[a]{Carlota Andres}

\affiliation[a]{CPHT, CNRS, Ecole polytechnique, Institut Polytechnique de Paris, \\
  91120 Palaiseau, France}

\emailAdd{carlota.andres@polytechnique.edu}

\abstract{These proceedings present a personal selection of emerging topics in jet and heavy-flavor physics in relativistic heavy-ion collisions, based on a plenary talk delivered at Hard Probes 2026 in Nashville (USA). }

\FullConference{13th International Conference on Hard and Electromagnetic Probes of High-Energy Nuclear Collisions (HardProbes2026)\\
21-26 June 2026\\
Nashville, Tennessee, USA\\}

\begin{document}
\maketitle

\section{Introduction}
\label{sec:intro}


Jets and heavy quarks are predominantly produced in hard scatterings at the earliest stages of a heavy-ion collision. Their large production scales, set by the jet transverse momentum or the heavy-quark mass, lead to short production times, making them sensitive to the subsequent evolution of the system. Through their interactions with the surrounding medium, these probes can provide insights into different stages of its evolution, including the pre-equilibrium phase, whose dynamics remains difficult to constrain using bulk observables. They also offer complementary ways of investigating the QGP: the multiscale nature of jets may provide access to its dynamics across a wide range of scales, while heavy quarks offer opportunities for a systematic description of their interactions with the medium using effective theories (EFTs). Realizing this potential requires well-motivated observables and increasingly controlled theoretical descriptions.

The discussion is organized around a selection of recent topics in both areas. These include novel observables in jet substructure, jet-medium interactions in the pre-equilibrium stages, and effective field theory (EFT) descriptions of open and hidden heavy-quark dynamics. Together, these topics illustrate different ways in which hard probes can deepen our understanding of the medium, from its microscopic properties to its dynamical evolution. This selection reflects my own perspective on recent advances in the field,  rather than an attempt at a comprehensive review. Many important contributions are therefore not covered, and their omission should not be taken as a judgment of their relative importance.

\section{Jets}
\label{sec:jets}

\subsection{Energy correlators}

One of the most exciting developments in heavy-ion physics in recent years has been the emergence of energy correlators as probes of the QGP. First proposed for this purpose only about four years ago \cite{Andres:2022ovj,Andres:2023xwr}, they have since moved rapidly from theoretical proposals to experimental measurements at both the Large Hadron Collider (LHC) and the Relativistic Heavy Ion Collider (RHIC), in p-p \cite{ALICE:2024dfl,STAR:2025jut,CMS:2024mlf,ALICE:2025igw,LHCb:2026bod}, p-Pb \cite{ALICE:2026giw,ALICE:2026zyx}, and Pb-Pb \cite{CMS:2025ydi} collisions. In the following, I will focus on the two-point energy correlator in the collinear limit (EEC).

 A key feature that sets energy correlators apart from other jet-substructure observables is their direct formulation in quantum field theory: beyond being defined as weighted cross sections, they can be expressed as expectation values of products of energy-flow operators \cite{Sveshnikov:1995vi,Korchemsky:1999kt, Hofman:2008ar}. This connection to fundamental field-theoretic objects opens the door to theoretical concepts and tools that are not applicable to more traditional substructure observables. In particular, the light-ray operator product expansion (OPE)~\cite{Hofman:2008ar} and celestial blocks~\cite{Chen:2022jhb,Chang:2022ryc} provide a powerful framework for  understanding their structure. The light-ray OPE allows us to expand the product of two energy-flow operators, separated by an angular distance $\theta$, as a series in $\theta$ with operator coefficients of fixed spin and increasing twist. The leading-twist contribution reproduces the measured classical $1/\theta$  scaling of the EEC in vacuum~\cite{Hofman:2008ar}. The picture becomes particularly interesting when moving from vacuum to nuclear  (p-A and A-A) collisions. As shown in~\cite{Andres:2024xvk}, applying the light-ray OPE to p-Pb and Pb-Pb collisions reveals the importance of higher-twist contributions, which are expected to be enhanced by the size of the nucleus or medium~\cite{Luo:1992fz,Luo1994Twist4NP}. This leads to an enhancement of the EEC relative to the vacuum baseline that grows  as $\theta^2$ at the classical level, in agreement with the CMS EEC measurement in Pb-Pb collisions~\cite{CMS:2025ydi,Andres:2024xvk}. This is an exciting example of how the QFT structure of energy correlators can lead to concrete predictions for experimentally measurable modifications of jet substructure in nuclear collisions. At the same time, while the QFT formulation provides a particularly powerful perspective, the cross-section definition remains important: the complexity of nuclear collisions calls for making use of every available theoretical tool. Indeed, energy correlators in nuclear collisions have attracted considerable theoretical activity, with studies based on both Monte Carlo event generators~\cite{Yang:2023dwc,Bossi:2024qho,Apolinario:2025vtx,Yang:2026bsk} and (semi-)analytical calculations~\cite{Andres:2022ovj,Andres:2023xwr,Andres:2024ksi,Barata:2023bhh,Ke:2025ibt,Bossi:2026mdz}. 
 
\begin{figure}[t]
   \centering
   \includegraphics[scale=0.46]{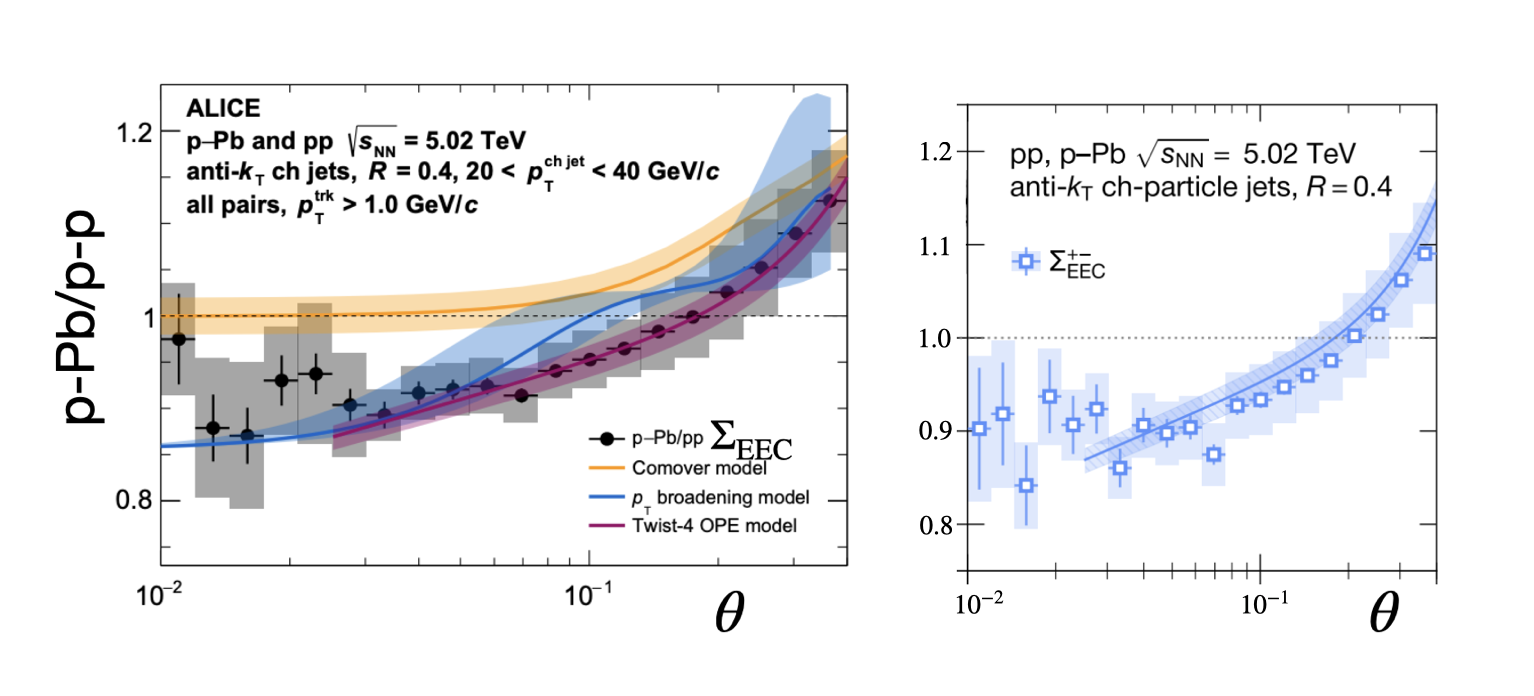}
   \caption{ Inclusive $\Sigma^{\rm EEC}$ (left) and charge-dependent $\Sigma^{\rm EEC^{+-}}$ (right) EEC ratios in p-Pb relative to p-p collisions for charged-particle jets with $p_T=20$–$40$ GeV/$c$. The left panel compares the measurement with the OPE (purple), transverse-momentum broadening (blue), and higher-twist + comovers (orange) calculations, while the right panel shows the comparison with the OPE + track functions calculation. Figure adapted from~\cite{ALICE:2026giw,ALICE:2026zyx}.}
  \label{fig:EEC_pPb}
 \end{figure}

Another particularly interesting result is the ALICE measurement of the EEC in p-Pb jets~\cite{ALICE:2026giw}, which shows an enhancement relative to p-p at large $\theta$ for jets with low $p_T$. To my knowledge, this represents the clearest modification of jet substructure observed in p-Pb collisions to date. Several theoretical explanations have been proposed, including twist-4 contributions in the OPE~\cite{Andres:2024xvk}, transverse-momentum broadening~\cite{Barata:2024wsu}, higher twist combined with comovers~\cite{Fu:2024pic}, and an SCET$_{\rm G}$ approach~\cite{Ke:2025ibt} (see the left panel of Fig.~\ref{fig:EEC_pPb}). Nevertheless, its origin remains an open question. More recently, ALICE has also measured charge-dependent EECs~\cite{Lee:2023npz} in p-Pb collisions~\cite{ALICE:2026zyx} (see also the STAR measurement in pp~\cite{STAR:2025jut}), and the OPE~\cite{Andres:2024xvk} combined with track functions~\cite{Chang:2013rca,Chang:2013iba} predicts the observed behavior  remarkably well, as shown in the right panel of Fig.~\ref{fig:EEC_pPb}.

Looking ahead, measurements of energy correlators with different energy weights (generalized detectors), as well as three-point correlators in A–A and p–A collisions, would be particularly valuable \cite{Gonzalez:2026aha,Aretz:inprep}. They would provide access to additional angular scales and, in the language of the OPE, probe the spin dependence of the relevant matrix elements, offering further handles to discriminate between different theoretical scenarios. Measurements in light-ion collision systems, such as oxygen-oxygen collisions, would be equally valuable, as we will see in the following.

\subsection{Jets in light-ion collisions}
\label{sec:lightions}
The 2025 light-ion run at the LHC has already delivered striking results. The first measurements show a strong suppression of high-$p_T$ hadrons in both oxygen-oxygen (O-O) and neon-neon (Ne-Ne) collisions~\cite{CMS:2025bta,CMS:2026qef,ALICE:2026zck} together with a dijet momentum imbalance that increases from peripheral to central collisions~\cite{ATLAS:2026fos}, as expected from jet quenching. Together with the semi-inclusive jet measurements in O-O performed by STAR~\cite{STAR:2026nfy} and the preliminary results on jet $R_{\rm AA}$ in O-O presented by sPHENIX at this conference~\cite{sPHENIX:2026jetOO},  these findings point towards energy loss in light-ion collisions. Particularly compelling is the centrality dependence of the dijet imbalance, which is difficult to explain through mechanisms other than quenching, and would make O--O collisions the smallest system in which jet quenching has been observed.  The implications extend beyond jet quenching: when considered together with other measurements, including the observation of elliptic  flow for both light and heavy flavors \cite{ALICE:2025luc,ATLAS:2025nnt,CMS:2025tga,DiCostanzo:2026qgo} and the sequential suppression of the $\Upsilon$ states~\cite{CMS:2026qbb}, the results strongly suggest that a QGP is formed in O-O collisions.

Although the measurements point toward energy loss, quantifying its magnitude remains challenging, as its extraction from single-inclusive suppression measurements is affected by the large uncertainties in nuclear parton distribution functions (nPDFs) for light nuclei~\cite{CMS:2025bta,ALICE:2026zck,sPHENIX:2026jetOO}. Indeed, the mass-number dependence of the gluon nPDF is almost unconstrained around $A=16$, owing to the lack of data in this mass-number regime~\cite{Paakkinen:2021jjp}.   Imposing new constraints on nPDFs  is therefore essential and can be achieved through dijet measurements in p–O collisions and electroweak boson production in p–O and O–O~\cite{Paakkinen:2021jjp}. A complementary strategy would be to focus on observables with reduced sensitivity to nPDFs~\cite{ATLAS:2026fos,Jonas:2026yoz,ALICE:2026zck}.

Moving beyond energy loss, one ingredient  still missing to complete the picture of quenching in light-ion collisions is the observation of jet-substructure modifications relative to p–p. Whether such modifications will be large enough to be observable in these smaller systems remains an open question.  However, both energy loss (and the associated selection bias) and the underlying-event background are also reduced compared with heavy-ion collisions. This creates an opportunity worth exploring: with sufficient statistics, the signal-to-background ratio might allow us to uncover effects that are more difficult to isolate in heavy-ion collisions~\cite{Kudinoor:2026wcs,Apolinario:2025fwd,Ke:2025ibt}. Furthermore, the lower background may make it possible to perform jet-substructure measurements using larger jet radii than those typically accessible in heavy-ion collisions, where reconstructed jets are usually measured with radii up to $R=0.4$. In particular, measuring energy correlators in light-ion collisions with $R=0.6$ or $R=0.8$ would be especially interesting~\cite{Kudinoor:2026wcs,Ke:2025ibt,Andres:2025yls}. In my opinion, O–O collisions offer a unique opportunity to attempt this program, as data are available at  both RHIC and the LHC. Furthermore, their shorter lifetime makes light-ion collisions a natural system in which to look for signatures of thermalization, which brings us to the next topic.

\subsection{Medium response}
\label{sec:wake}

As a jet propagates through the QGP,  it deposits energy and momentum into the medium, giving rise to an excess of soft particles along the jet direction referred to as (the positive) wake~\cite{Cao:2020wlm}. Although these particles originate from the medium rather than from the jet fragmentation itself, they remain correlated with the jet and are therefore clustered into the reconstructed jet. As a result, a reconstructed jet contains not only the products of its own fragmentation but also particles generated by this medium response. Separating these two contributions and identifying an unambiguous signature of the wake has been a long-standing challenge.
An especially important aspect of this response follows from momentum conservation: the excess of soft particles in the jet direction is accompanied by a depletion in the opposite direction, known as the negative wake. CMS has observed such a  depletion in Pb-Pb collisions, both in Z–hadron correlations on the side of the Z boson~\cite{CMS:2025dua} and, more recently, in dijet events with a pseudorapidity gap between the two jets~\cite{CMS:2026mur}. The gap prevents the positive and negative wakes from overlapping, making the latter observable (see~\cite{Pablos:2019ngg,Yang:2025dqu}).

While these measurements do not yet provide a direct observation of the wake inside the jet cone, they represent an important step towards understanding how the medium response contributes to jet observables.  More broadly, they open an exciting connection to the question of thermalization of the QGP, since the evolution of the wake is expected to be governed by the same underlying dynamics that drive the thermalization of the bulk medium in the early stages of the collision. In this sense, studying the wake could provide a new handle on how the QGP approaches local thermal equilibrium. The first theoretical studies of jet thermalization in kinetic theory have begun to explore this question~\cite{Mehtar-Tani:2022zwf,Soudi:2025lei}.

\subsection{Jet-medium interactions in the pre-equilibrium stages}
\label{sec:timescales}

A complementary way to access the thermalization process may be to use jets as time-differential probes of the system evolution, exploiting their sensitivity to its different stages, including the pre-hydrodynamic phases \cite{Andres:2019eus,Zigic:2019sth}.  Consider two vacuum parton showers ordered in different variables, such as emission angle or inverse formation time.  Although these showers are equivalent at double-logarithmic accuracy, they are not identical splitting by splitting and therefore experience different amounts of energy loss when embedded in an evolving medium~\cite{Andres:2024egc}. In this way, jets can encode information not only about the properties of the medium, but also about when they interact with it.
This time sensitivity motivates reclustering jets according to formation time. The $\tau$ algorithm~\cite{Apolinario:2020uvt,Apolinario:2024hsm} makes it possible to select splittings with short or long formation times, opening the possibility of using jets as chronometers of the QGP. An ambitious goal would be to isolate splittings that take place during the pre-equilibrium stages. Reaching this goal, however, requires a better understanding of how jets interact with the medium before it enters the hydrodynamic regime.

This question has prompted a growing body of theoretical work on jet interactions in the pre-equilibrium stages. Several recent classical calculations show that high-energy partons undergo substantial transverse-momentum broadening in the Glasma~\cite{Ipp:2020mjc,Ipp:2020nfu,Carrington:2021dvw,Carrington:2022bnv}. This broadening is anisotropic, being larger along the beam axis. Despite their different approaches, the calculations yield consistent results for the shape and overall magnitude of the broadening coefficient $\hat q$, which is controlled mainly by the saturation scale $Q_s$ and depends only weakly on other parameters. A similar picture emerges from calculations in QCD effective kinetic theory (EKT)~\cite{Boguslavski:2024ezg,Boguslavski:2024jwr}: broadening is likewise large and appears to connect smoothly to the Glasma results.

While these calculations establish that substantial transverse-momentum broadening occurs in the pre-equilibrium stages, they do not yet tell us whether jets also lose a significant amount of energy during this short period of time. The Glasma calculations are classical, and the obtained transport coefficient $\hat q$ characterizes transverse-momentum broadening only. A large $\hat q$ does not necessarily imply large energy loss, since the connection between the calculated broadening and medium-induced radiation is not yet established. A recent calculation goes beyond the classical treatment by evolving the quark wave function in the Glasma fields within a light-front Hamiltonian framework~\cite{Avramescu:2026fgl}. Although so far only quantum broadening has been computed, the framework can in principle also incorporate gluon radiation, opening the possibility of calculating energy loss for the first time in realistic Glasma fields. Even once this is achieved, a further challenge will be to identify which jet observables retain sensitivity to energy loss in the pre-equilibrium stages and how to disentangle such effects from those arising during the hydrodynamic phase.

\section{Heavy flavor}
\label{sec:hf}

\subsection{Heavy-quark-medium interactions in the pre-equilibrium stages}
\label{sec:glasmahq}

At low $p_T$, heavy quarks offer a complementary and theoretically cleaner probe of the pre-hydrodynamic medium. In this $p_T$ regime, radiation can be neglected and thus heavy-quark interactions with the medium are considerably easier to describe than those of jets, for which medium-induced radiation remains an important complication, particularly in the Glasma. Several classical approaches have  found substantial momentum broadening of heavy quarks in the Glasma. These include calculations based on real-time solutions of the Yang–Mills equations coupled to Wong's equations~\cite{Khowal:2021zoo,Avramescu:2023qvv}, as well as approaches based on a Fokker–Planck equation with collision terms computed from Glasma field correlators~\cite{Carrington:2020sww,Carrington:2022bnv}.

\begin{figure}
   \centering
   \includegraphics[scale=0.79]{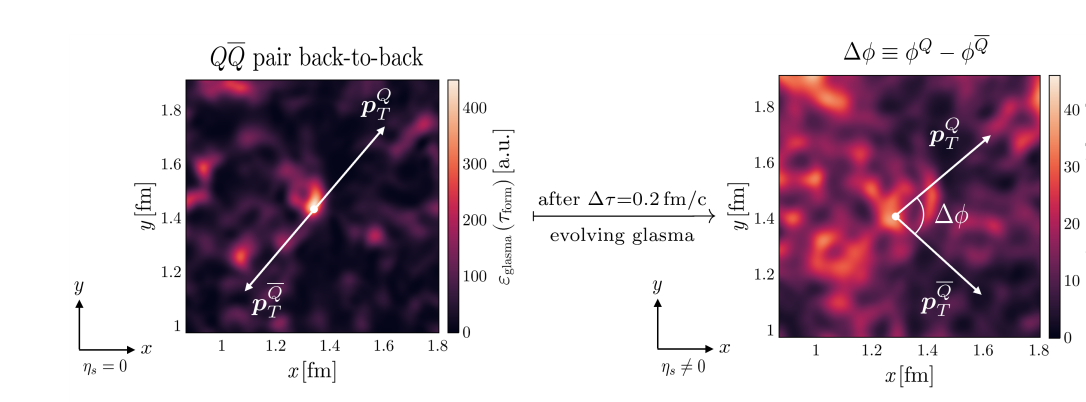}
   \caption{ Schematic representation of a $Q\bar{Q}$ pair evolving in the glasma background field. The quarks are initially back-to-back. After $\Delta\tau$, interactions with the evolving glasma modify their relative azimuthal angle $\Delta\phi$ and pseudorapidity $\Delta\eta$, which were initially peaked at $(\pi,0)$. Figure adapted from \cite{Avramescu:2024poa}.}
  \label{fig:HQ_deco}
 \end{figure}

The next question is whether this early-time broadening leaves a measurable imprint on heavy-flavor observables. So far, the D-meson nuclear modification factor and elliptic flow have been found to be only mildly sensitive to the Glasma stage~\cite{Sun:2019fud, Avramescu:2024poa,Singh:2025duj}.   A more promising avenue is offered by azimuthal correlations of heavy-quark pairs~\cite{Avramescu:2024poa,Avramescu:2024xts}. A $c\bar c$ pair produced back-to-back progressively loses its azimuthal correlation through interactions with the Glasma fields (see Fig.~\ref{fig:HQ_deco}), with a decorrelation comparable in magnitude to that accumulated during the subsequent hydrodynamic evolution~\cite{Nahrgang:2013saa,Cao:2015cba,Sun:2019fud}. In this respect, phenomenological studies are arguably more advanced for heavy flavor than for jets, although disentangling the Glasma and hydrodynamic contributions remains challenging. This makes $D\bar D$ correlations a particularly interesting target for the high-luminosity era of Pb–Pb collisions.

The picture extends beyond classical Glasma calculations. Heavy-quark broadening has also been computed in EKT~\cite{Boguslavski:2020tqz,Boguslavski:2023fdm,Du:2023izb,Danhoni:2026wjl}, where it is also found to be relatively large. However, unlike for jets, the Glasma and EKT results do not connect as smoothly, leaving open the question of how to consistently describe heavy-quark interactions across the different stages of the evolution.

\subsection{Heavy-quark transport beyond the Gaussian approximation}
\label{sec:nongaussian}
Once the medium enters the hydrodynamic phase, open heavy-flavor phenomenology is largely described using Fokker–Planck or Langevin approaches. These frameworks typically impose the Einstein relation between the drag and diffusion coefficients to ensure that heavy quarks thermalize at sufficiently large times. For a heavy quark at rest, this relation follows from the fluctuation-dissipation theorem. At finite velocity, however, as is particularly relevant for charm quarks, no such theorem guarantees its validity. Indeed, the relation is violated in some QFTs, both at weak coupling beyond leading-logarithmic accuracy~\cite{Moore:2004tg}, and at strong coupling in $\mathcal N=4$ supersymmetric Yang-Mills theory~\cite{Gubser:2006nz}.

This raises a more general question: what governs the equilibration of a moving heavy quark when the Einstein relation no longer holds? A universal relation for the probability distribution of momentum transfers between a moving heavy quark and the medium has recently been derived~\cite{Rajagopal:2025rxr}.  Remarkably, it holds in any QFT and at any coupling, provided the heavy-quark mass is much larger than the medium temperature. While it reduces to the familiar Einstein relation when momentum transfers are Gaussian, in the general case it constrains the full hierarchy of moments of the momentum-transfer distribution. Equilibration is therefore governed by their interplay, rather than by the first two moments alone. This can have important consequences for heavy-quark equilibration, particularly for charm, where non-Gaussian effects can significantly modify the thermalization dynamics~\cite{Rajagopal:2025rxr,Rajagopal:2025ukd,DuPlessis:2026pyr}. In particular, both the thermalization time and the spatial diffusion coefficient differ from their values obtained in the Gaussian limit. Similar effects have also been observed phenomenologically through comparisons between Langevin and Boltzmann approaches~\cite{Das:2013kea,Scardina:2017ipo}.

\subsection{EFT for heavy-quark dynamics}
\label{sec:fluidEFT}

A complementary perspective on open heavy-flavor dynamics is to describe how heavy quarks interact with the QGP in terms of its collective degrees of freedom. 
A recently developed EFT provides a framework for this approach, with the potential to interpolate between the weakly and strongly coupled regimes~\cite{Kirchner:2025xdb}. The underlying idea is that, if the QGP behaves as a fluid, heavy quarks can be coupled directly to an EFT of fluid dynamics. In its current formulation, the QGP is described as a non-dissipative relativistic fluid, parametrized by scalar fields that track the motion of individual fluid elements~\cite{Endlich:2010hf,Dubovsky:2011sj}. The excitations around equilibrium are sound waves, or phonons, and heavy quarks therefore interact with the collective excitations of the fluid, rather than directly with its microscopic constituents. This description leads to a particularly intuitive picture of heavy-quark energy loss. A moving heavy quark can absorb phonons and, when moving faster than the speed of sound, emit them.  The emitted phonons can generate a characteristic Mach cone in the medium. An especially appealing feature of this framework is its potential to incorporate dissipative effects. This would make it possible to describe the diffusion wake generated by heavy quarks.

\subsection{Quarkonia as open quantum systems}
\label{sec:quarkonia}

While EFT approaches to open heavy-flavor dynamics have only recently begun to develop, their application to hidden heavy flavor has a longer history. Quarkonia are characterized by a clear hierarchy of scales: the heavy-quark mass, the inverse size of the bound state, and its binding energy are well separated, allowing for a systematic description using EFTs. Placing quarkonia in a medium, however, raises two additional questions: how does the temperature fit into this hierarchy, and how can the intrinsic dynamics of the bound state be disentangled from that of the QGP?  Both questions can be addressed within the open quantum system (OQS) framework combined with EFTs~\cite{Akamatsu:2011se,Akamatsu:2014qsa}. The starting point is to consider the heavy quark-antiquark pair and the QGP together as a closed system. By evolving this system in time and tracing over the QGP degrees of freedom, one obtains a master equation for the pair, which takes the Lindblad form in the Markovian limit.

The precise form of this equation, and therefore the resulting dynamics, depends on the hierarchy between the temperature and the binding energy. When the temperature is large compared with the binding energy, one enters the quantum Brownian limit, in which the QGP resolves the quark and the antiquark individually, so that each interacts with the medium almost independently.  The resulting dynamics is driven by diffusion and decoherence~\cite{Miura:2022arv,Delorme:2024rdo,Brambilla:2024tqg}. Conversely, when the temperature is small compared with the binding energy, one enters the quantum optical limit, in which the medium interacts with the pair as a whole, and its effect is described in terms of transitions between the levels of the quarkonium spectrum~\cite{Yao:2018nmy}. These two limits are therefore expected to describe different stages of the QGP evolution, with the Brownian limit being better suited to the early stages and the optical limit to the later ones.
A natural next step is to connect these two regimes within a common framework. This is the goal of a recently derived universal Lindblad equation~\cite{Hammou:2026yqh}, building on~\cite{Nathan:2020qck}. It reduces to either limit when the corresponding hierarchy is imposed, providing an important step towards a unified quantum description of quarkonium dynamics in the QGP. Its first applications to charmonium spectral functions, for example, show the rapid melting of the P-wave state as the temperature increases~\cite{Hammou:2026yqh}.

\begin{figure}
   \centering
   \includegraphics[scale=0.43]{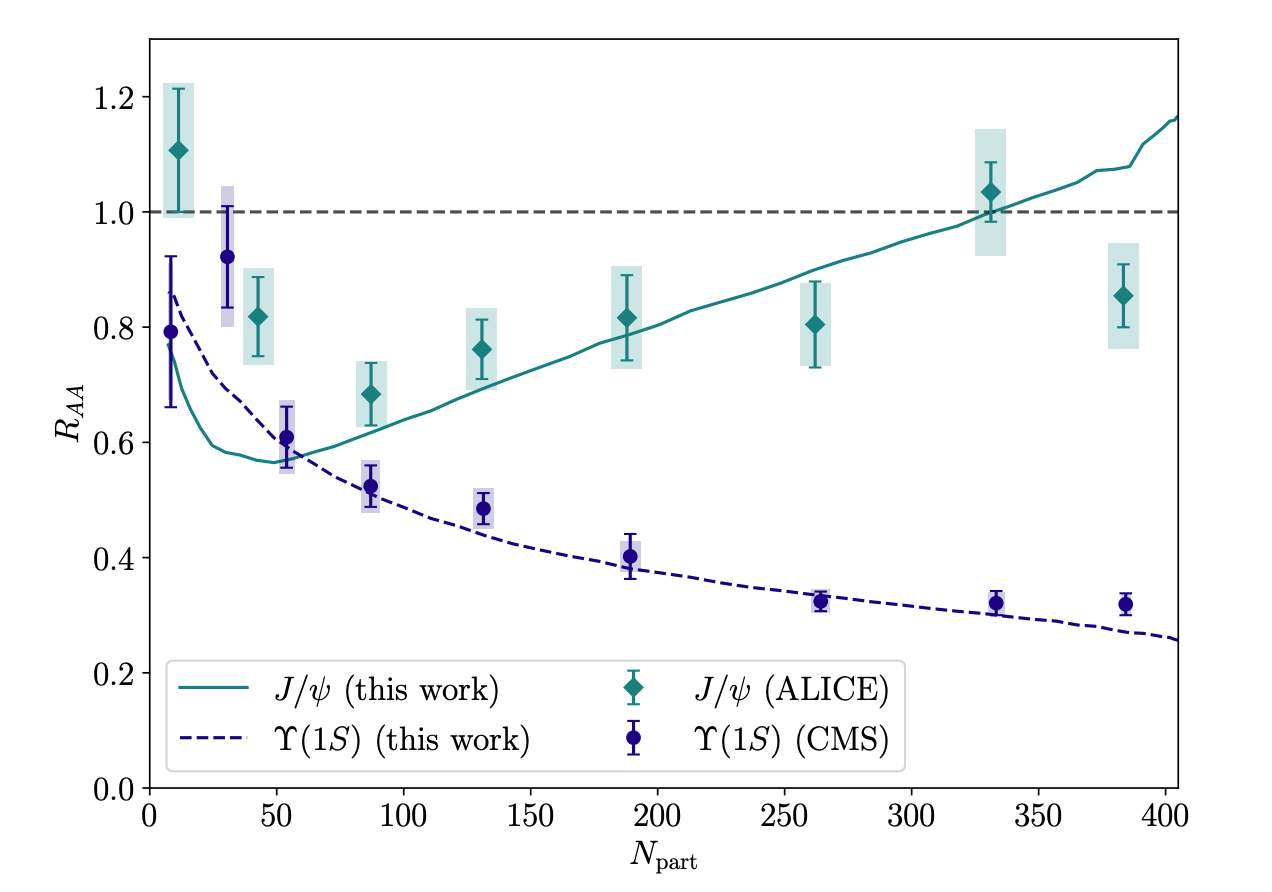}
   \caption{Solid lines: Nuclear modification factor $R_{AA}$ of the $J/\psi$ and $\Upsilon(1S)$ in Pb-Pb collisions at $\sqrt{s_{\rm NN}} = 5.02$ TeV, calculated within the framework of~\cite{Armesto:2026fit}. Experimental data from ALICE~\cite{ALICE:2019pid} and CMS~\cite{CMS:2017ycw,CMS:2017uuv} are shown for comparison. Figure taken from~\cite{Armesto:2026fit}.}
  \label{fig:RAA_quarkonia}
 \end{figure}

A further challenge for OQS approaches is to go beyond the evolution of a single heavy quark-antiquark pair and account for quarkonium recombination. A recent study addresses this issue by deriving a recombination equation from a Lindblad equation with a given complex potential~\cite{Armesto:2026fit}. The resulting rate equation has the same form as those used for many years in phenomenology, but with an important difference: the same complex potential determines both the thermal width and the recombination dynamics. Once the potential is fixed, no free parameters remain, and the recombination rate no longer needs to be constrained through detailed balance. Its results are in agreement with $J/\psi$ and $\Upsilon(1S)$ suppression data, as shown in Fig.~\ref{fig:RAA_quarkonia}.

The interplay between open and hidden heavy flavor is also being explored in transport approaches. A recent framework couples the two in a strongly coupled QGP~\cite{Fu:2026ous,Du:2022uvj}: charm-quark diffusion, described by Langevin dynamics, is coupled to a Boltzmann equation for charmonium dissociation and regeneration.

\section{Conclusions}
\label{sec:conclusions}

Taken together, the developments discussed here reflect progress on two closely connected fronts: improving our theoretical understanding of how hard probes interact with the QGP, and expanding the questions we can address with them. On the theory side, considerable progress has been made in developing QFT-rooted frameworks to describe different aspects of these interactions. These include the light-ray OPE for energy correlators, universal equilibration conditions for heavy quarks, EFT descriptions of heavy-quark dynamics in the QGP, and open-quantum-system approaches to quarkonia. At the same time, the questions we are asking with hard probes are becoming increasingly ambitious. Beyond extracting transport coefficients, we aim to use hard probes as multiscale tools to investigate the microscopic dynamics of the QGP and identify possible signatures of its pre-hydrodynamic stages. The theoretical developments in this direction over the past five years have been remarkable.

On the experimental side, measurements that seemed out of reach just five years ago are now becoming a reality. These include the observation of the negative wake and a growing range of high-precision jet-substructure measurements, such as energy correlators in Pb–Pb and p–Pb collisions. Most of these results, however, are based on LHC Run 2 data, making the prospects offered by the higher precision of Run 3, together with the opportunities offered by the new sPHENIX detector at RHIC, particularly exciting. Light-ion collisions have also opened up a new laboratory that we have only begun to explore.  At both RHIC and the LHC, the first analyses of hard probes in O–O collisions are now emerging. Having O–O measurements at both facilities, at two different center-of-mass energies, will provide a particularly valuable opportunity to explore QGP signatures in small collision systems and gain complementary insights into the underlying physics. The coming years will be crucial for sharpening the questions we can address and setting ambitious goals for Run 4.

\acknowledgments
I am grateful to Liliana Apolin\'ario, N\'estor Armesto, Dana Avramescu, Cristian Baldenegro, Hannah Bossi, Elena G.~Ferreiro, Andreas Kirchner, V\'ictor L\'opez-Pardo, Cyrille Marquet, Aleksas Mazeliauskas, Krishna Rajagopal, Bruno Scheihing-Hitschfeld and Xiaojun Yao for helpful discussions. This work is funded by the
European Union (ERC, QGPthroughEECs, grant agreement No. 101164102). Views and opinions expressed are however those of the authors only and do not necessarily reflect those of the
European Union or the European Research Council. Neither the European Union nor the granting authority can be held responsible for them.

\bibliographystyle{JHEP}
\bibliography{refs}

\end{document}